\documentclass[letterpaper,twocolumn,10pt]{article}
\usepackage{usenix2019_v3}

\usepackage{xurl}

\usepackage{svg}

\usepackage{csquotes}

\usepackage{orcidlink}

\usepackage{booktabs}
\usepackage{listings}
\newcommand{\todo}[1]{}
\newcommand{\todoanswer}[1]{}
\newcommand{\changedtext}[1]{#1}

\newcommand{\anonymized}[2]{#2}

\begin{document}

\date{}

\title{\Large \bf Charge It to My Neighbor: A Relay Attack on ISO 15118 Plug and Charge Payment}

\author{
{\rm Jakob L\"{o}w \orcidlink{0009-0006-7088-8684}} \\
Technische Hochschule Ingolstadt
\and
{\rm Vishwa Vasu \orcidlink{0009-0009-5996-5601}}\\
Technische Hochschule Ingolstadt
\and
{\rm Thomas Hutzelmann \orcidlink{0000-0002-2871-3905}}\\
Technische Hochschule Ingolstadt
\and
{\rm Hans-Joachim Hof \orcidlink{0000-0002-6930-9271}}\\
Technische Hochschule Ingolstadt
} 

\maketitle

\begin{abstract}
ISO 15118, the leading standard for DC fast charging in Europe, includes a plug-and-charge mechanism that allows electric vehicles to handle payment automatically via contract certificates.
We present a novel relay attack against this mechanism: an attacker builds a fake charging station, plugs it into a victim's vehicle, and relays the cryptographic authentication to a real charging station --- charging the attacker's vehicle while billing the victim.
The attack exploits the absence of station-identifying information in the plug-and-charge signature, combined with weaknesses in how ISO 15118 handles TLS certificates.
We provide a proof-of-concept implementation demonstrating the full attack chain and discuss possible mitigations and alternatives.
As plug-and-charge adoption grows, addressing this vulnerability is critical before it becomes widely exploitable.
\end{abstract}


\section{Introduction}
Electric vehicles are gaining in popularity.
Especially in countries with corresponding incentives to buy electric over combustion engines, like China and Norway, they reach nearly 100\% market share in newly registered vehicles \cite{noauthor_openevcharts_2025}.
With the increasing number of electric vehicles on the road also comes a growing demand for fast charging stations, allowing electric vehicles to quickly recharge before continuing their trip.

In Europe, the leading standard for charging communication between an electric vehicle and a charging station is ISO 15118 \cite{isoiec_isoiec_2012, isoiec_isoiec_2012-1, isoiec_isoiec_2012_20}.
This standard not only describes the required communication between the two parties but also includes additional features, such as plug and charge payment, allowing the user to handle payment directly after plugging in a vehicle without further user interaction.
A recent study found that plug and charge is the most preferred yet least used payment method among electric vehicle drivers, while mobile app and RFID card based payment are the most used but least preferred \cite{uscale_emobility_2025}.
This significant discrepancy suggests that plug and charge adoption will continue to grow.
Handling payment directly within the charging communication, however, also increases the attack surface and, especially, the reward potential that attackers could have when attacking charging communication.

In this paper, we present a novel relay attack against the plug and charge payment mechanism of ISO 15118.
An attacker builds a fake charging station, plugs it into a victim's vehicle, and relays the cryptographic payment authentication to a real charging station --- charging the attacker's vehicle while billing the victim.
The attack exploits the absence of station-identifying information in the plug and charge signature, combined with weaknesses in how ISO 15118 handles TLS certificates.
We provide a proof-of-concept implementation demonstrating the full attack chain and discuss possible mitigations, including an alternative out-of-band payment approach that avoids the vulnerability entirely.

\changedtext{While, to the best of our knowledge, the relay attack itself has not been described before, our proof-of-concept implementation builds on shortcomings in charging communication security identified in related work \cite{szakaly_current_2025, szakaly_short_2025, eder_charging_2025, szakaly_dception_2026}.}
\changedtext{Conti et al.\ \cite{conti_evexchange_2022} have previously demonstrated an attack against plug and charge payment; however, their approach relies on opaque forwarding, which introduces several limitations discussed later.}

Section \ref{sec:background} provides background on the plug and charge mechanism and TLS certificate handling in ISO 15118.
Section \ref{sec:related-work} discusses related work on charging communication security.
Section \ref{sec:adversary-model} introduces the adversary model used for evaluating the vulnerability.
Section \ref{sec:pnc-vuln} presents the plug and charge relay vulnerability.
Section \ref{sec:poc-impl} describes our proof-of-concept implementation.
Finally, Section \ref{sec:mitigations} discusses mitigations and an alternative payment approach.

\todo{[R13B] The Introduction would benefit from a stronger positioning with respect to related work, including additional references to prior studies in this space.}
\todoanswer{We extended the intro, mentioning the relation to previous work.}


\section{Background} \label{sec:background}
\todo{[R13B] Section 2 would be clearer with the inclusion of a figure or diagram illustrating the protocol workflow.}
\todoanswer{added a new figure similar to the relay attack explanation, but showing the "normal" communication flow.}
ISO 15118 \cite{isoiec_isoiec_2012, isoiec_isoiec_2012-1, isoiec_isoiec_2012_20} defines the communication protocol between electric vehicles and charging stations for DC fast charging in Europe.
The standard uses powerline communication over the charging cable, with IPv6 for network addressing and optionally TLS for session encryption.
Before the main charging communication begins, several initialization steps take place: low-level signaling per IEC 61851 \cite{iec_iec_2010}, powerline network formation via the Signal Level Attenuation Characterization (SLAC) protocol, IPv6 address assignment, \changedtext{and TCP connection parameter exchange using the Service Discovery Protocol (SDP)}.
Previous work has identified vulnerabilities in many of these initialization steps \cite{eder_charging_2025, szakaly_current_2025, kohler_brokenwire_2023, dudek-v2g-2019}.
For this paper, the relevant aspects of ISO 15118 are the plug and charge payment mechanism and the use of TLS certificates, both described below.

\subsection{Plug and Charge} \label{sec:plugandcharge}
\todo{[R13C] For readers who are not familiar with the topic, some aspects could be explained in more detail. In addition, abbreviations are used without spelling out the full term, such as SDP in Section 2.1.}
\todoanswer{SDP is now spelled out on first use.}
After the SDP request and response described in the previous section, the two communication parties establish a TCP connection.
When TLS is supported by both parties, a standard TLS handshake follows.
Within this established connection, the main charging communication is performed.
The vehicle sends a \texttt{ServiceDiscoveryReq} to which the charging station responds with a \texttt{ServiceDiscoveryRes}.
This response includes a list of supported payment methods.
As of today, mostly the \texttt{ExternalPayment} method is used \cite{szakaly_current_2025, low_fast_2024}.
With this method, the user has to handle payment directly with the charging station, for example, using a credit card.

Alternatively, the charging station may offer \texttt{ContractCertificate} payment.
\changedtext{If the vehicle selects this payment method in its \texttt{PaymentServiceSelectionReq}, it follows up with its certificate in a \texttt{PaymentDetailsReq}}.
The charging station responds with a \texttt{PaymentDetailsRes}, including a unique challenge consisting of random bytes.
The vehicle has to sign the challenge bytes using the private key belonging to the given certificate, sending the signature in a \texttt{AuthorizationReq}.
This way, the vehicle proves to the charging station that it has ownership of the private key belonging to the contract certificate presented in the previous step.
\changedtext{Figure \ref{fig:pnc-flow} illustrates this communication flow, when plug and charge payment is used.}

\begin{figure}[ht]
    \centering
    \includesvg[width=0.9\linewidth]{graphics/iso15118-pnc.drawio.svg}
    \caption{Plug and Charge Flow}
    \label{fig:pnc-flow}
\end{figure}

According to the ISO 15118 standard \cite{isoiec_isoiec_2012}, plug and charge shall only be used when the connection is encrypted using TLS.
For establishing a TLS connection with ISO 15118-2 \cite{isoiec_isoiec_2012}, only the server needs to authenticate itself.
With the newer ISO 15118-20 \cite{isoiec_isoiec_2012_20}, mutual authentication is required for establishing an encrypted connection.
With ISO 15118, there are two separate public key infrastructures (PKIs).
One provides certificates to charging stations and vehicles for establishing TLS communication.
While ISO 15118-2 requires only the charging station to possess a certificate, the newer ISO 15118-20 demands that both sides authenticate using a valid certificate obtained from the first PKI.
The second PKI is responsible for handing out contract certificates used for payment.
On a technological level, contract certificates used for payment are identical to certificates used for session authentication.
While a vehicle typically has only one fixed certificate for session authentication, this separation enables it to have multiple, possibly user-configurable contract certificates.


\subsection{TLS Certificate Handling in ISO 15118} \label{sec:tls-problems}
For securing the main charging communication, the ISO 15118 standard uses Transport Layer Security.
While TLS is a well-established protocol for securing and authenticating sessions, the way ISO 15118 employs it leaves gaps that are critical enablers for the relay attack presented in this paper.

TLS uses certificate-based authentication with a public key infrastructure for creating and signing certificates.
On the web, where TLS is commonly used, this means there are a number of root certificate authorities.
Website operators can acquire a certificate for their website from any of the commonly trusted certificate authorities.
Web browsers usually bundle a large number of certificate authority root certificates and update them regularly.
In the vehicle ecosystem, where storage space and internet connectivity are sparse, managing large certificate stores and especially updating them is significantly harder.
As of today, there are multiple certificate authorities for charging communication; however, it is not clear to the public which OEMs ship their vehicles with trust for which authorities.
\todo{[R13A] While it is true that "it is not clear to the public which OEMs ship their vehicles with trust for which authorities", in the sense that it is not standardized, there are some de facto standards in this field.}
\todoanswer{Based on our experience most CPOs (in Europe) use Hubject as their root certificate authority. Maybe this is the de-facto standard you are referring to. In order to prove this "to the public", field tests would have to be carried out with various charging stations, which is out of scope for this paper (but we are working on it).}

The most significant problem for the attack presented in this paper is the lack of verifiable charging station identification.
When on the web connecting to a website \texttt{example.com} using TLS, the web browser ensures the presented certificate was issued to \texttt{example.com}.
For charging communication, the ISO 15118 standard requires the charging station ID to be encoded as the certificate identity.
While this identity is unique per charging station, there is no way for the vehicle to check if the provided certificate belongs to the charging station currently connected to \cite{szakaly_current_2025}.
A potential attacker could acquire one certificate from any charging station and use this certificate to impersonate any other charging station.
While certificate compromise can happen on the web as well, the impact is different: \changedtext{on the web}, when a certificate for \texttt{example.com} is known to an attacker, the attacker can only impersonate this one website.
\todo{[R13A] Page 2, ""different: On the web". The O should be lowercase.}
\todoanswer{fixed}
With ISO 15118, when a certificate for one charging station is compromised, an attacker can impersonate any charging station, including stations from other vendors, operators or in other countries.
This creates a weakest-link problem: the security of the entire ecosystem is bounded by the least secure charging station.
Real-world evidence suggests that obtaining certificates from charging stations is feasible.
Even Alpitronic, the largest DC fast charging hardware vendor in Europe, has been affected by vulnerabilities that could enable certificate extraction.
In 2024, it was discovered that the configuration panel of Alpitronic Hyperchargers is accessible from the powerline communication network, meaning any connected vehicle can reach the charging station's administrative interface \cite{vanbeijnum_hacking_2024}.
Additionally, a CISA advisory revealed that Alpitronic Hyperchargers shipped with default credentials for this administrative interface \cite{cisa_alpitronic_2024} (CVE-2024-4622).
While neither vulnerability has been publicly demonstrated to lead to certificate extraction, they illustrate the kind of access an attacker could leverage, and the fact that they affect the market-leading vendor underscores the weakest-link concern.

\section{Related Work} \label{sec:related-work}
With the increasing adoption of electric vehicles, more and more cybersecurity researchers have uncovered vulnerabilities in ISO 15118.
Most related work has focused on the initialization phases of charging communication, such as vulnerabilities in powerline communication \cite{kohler_brokenwire_2023, eder_charging_2025, dudek-homeplugav-2015}, the SLAC process \cite{eder_charging_2025, dudek-v2g-2019}, and service discovery \cite{multin_iso_2018, bao_threat_2018}.
While these lower-layer vulnerabilities can facilitate man-in-the-middle positions that complement our attack (see Section \ref{sec:poc-real-world}), the relay vulnerability we present operates independently at the application layer.

The most closely related work is EVExchange by Conti et al. \cite{conti_evexchange_2022}, which presents a relay attack on ISO 15118 charging communication.
In EVExchange, an attacker places two malicious devices at two adjacent charging stations managed by the same backend infrastructure.
These devices forward the entire network traffic between the two EVSEs, effectively swapping which vehicle communicates with which charging station.
Since the relay operates at the network layer, it does not need to decrypt TLS-encrypted traffic; instead, it opaquely forwards all packets.
The attacker can then control when to stop each vehicle's charge, making the victim pay for the attacker's energy consumption.
EVExchange requires both the attacker and victim vehicles to be physically present at the same charging location, as the malicious devices must be installed between the vehicles and adjacent EVSEs.
As a countermeasure, Conti et al. propose a distance bounding protocol that measures round-trip times of rapid packet exchanges to detect the additional latency introduced by the relay.

Because EVExchange relays the entire charging session at the network layer, the charging parameters sent by each vehicle must be consistent with the physical charging station it is actually connected to.
In particular, during DC charging, the vehicle continuously reports its present battery voltage in \texttt{CurrentDemandReq} messages.
Both vehicles must therefore have the same battery voltage for the relayed session to proceed without errors.
Even when two vehicles start at the same voltage, the attacker's vehicle will see its battery voltage increase as it charges, causing a growing mismatch with the victim's reported values.
The only way to avoid this inconsistency would be to charge both vehicles simultaneously, which eliminates the attacker's economic gain.

Table \ref{tab:evexchange-comparison} summarizes the key differences between EVExchange and the vulnerability presented in this paper.
In contrast to EVExchange, the plug and charge relay vulnerability presented in this paper (Section \ref{sec:pnc-vuln}) operates at the application layer, specifically targeting the plug and charge authentication mechanism.
Rather than forwarding entire network flows, our attack selectively relays the cryptographic messages used for payment authentication: the contract certificate, the challenge, and the signed response.
This allows the attacker vehicle and the fake charging station to be at entirely different physical locations, communicating over a cellular network, similar to keyless go relay attacks on vehicles.
Furthermore, while EVExchange requires tampering with existing charging infrastructure, our attack only requires the attacker to build a portable fake charging station that is plugged into the victim's vehicle.

\begin{table*}[t]
\centering
\caption{Comparison of EVExchange and this work}
\label{tab:evexchange-comparison}
\begin{tabular}{lll}
\toprule
\textbf{Property} & \textbf{EVExchange \cite{conti_evexchange_2022}} & \textbf{This paper} \\
\midrule
Relay layer          & Network (IP)             & Application (ISO 15118) \\
Relayed data         & All network traffic      & Payment authentication messages \\
Vehicle Requirements & Same battery voltage     & None \\
Actual Charging      & Both vehicles            & Only attacker vehicle \\
TLS broken           & No (opaque forwarding)   & No (separate sessions) \\
Countermeasure       & Distance bounding        & Context binding / backend \\
\bottomrule
\end{tabular}
\end{table*}


\section{Adversary Model} \label{sec:adversary-model}
Bao et al. \cite{bao_threat_2018} have published an extensive threat analysis regarding ISO 15118, including the modeling of various adversaries.
We utilize their adversary models for evaluating the impact of the described vulnerabilities and their potential exploitation by attackers.
Of the adversaries defined by Bao et al., two are particularly relevant to the plug and charge relay vulnerability presented in this paper.
Additionally, we introduce a third adversary, the \textit{equipment supplier}, who is not defined by Bao et al. but is relevant for assessing the real-world scalability of the attack.

The \textit{freeloader} aims to charge their own vehicle for free by making someone else pay for the charged energy.
The freeloader has basic technical knowledge of electric vehicle charging and physical access to publicly accessible vehicles and charging stations.
However, the freeloader does not have the capability to develop custom hardware or software and instead relies on readily available tools or equipment obtained from third parties.
Using the relay attack described in Section \ref{sec:pnc-vuln}, a freeloader could build or acquire a fake charging station, plug it into a victim's vehicle, and relay the payment authentication to a real charging station where the freeloader's vehicle is charging.

The \textit{contract-sharer} aims to share one charging contract with a large number of other users, possibly abusing flat charging fees.
Similar to the freeloader, the contract-sharer has basic technical knowledge and physical access to charging infrastructure, but does not necessarily possess advanced hardware or software development skills.
The contract-sharer's distinguishing characteristic is the organizational capability to coordinate multiple users and distribute relay equipment or access credentials among them.
By relaying the plug and charge authentication from a single contracted vehicle, a contract-sharer could enable multiple unauthorized vehicles to charge under the same contract.

The \textit{equipment supplier} manufactures and sells ready-to-use relay equipment to customers such as freeloaders and contract-sharers.
Unlike the other two adversaries, the equipment supplier possesses advanced technical knowledge of charging communication protocols, powerline communication, and embedded systems development.
The equipment supplier is capable of developing custom hardware and software, such as portable fake charging stations and man-in-the-middle devices that combine known powerline communication vulnerabilities \cite{eder_charging_2025, dudek-v2g-2019, kohler_brokenwire_2023} with the plug and charge relay vulnerability described in Section \ref{sec:pnc-vuln}.
The equipment supplier's motivation is financial profit from selling attack equipment rather than directly exploiting the vulnerability.
Similar to the market for keyless go relay devices used in vehicle theft, the existence of an equipment supplier would significantly lower the barrier to entry for freeloaders and contract-sharers, as they would no longer need any technical expertise to carry out the attack.


\section{Plug and Charge Relay Vulnerability} \label{sec:pnc-vuln}
As described in Section \ref{sec:plugandcharge}, ISO 15118 supports contract-based payment, also known as plug and charge.
With this method, the vehicle transmits its contract certificate to the charging station, which responds with a random challenge.
The vehicle then signs this challenge using the private key belonging to the contract certificate, proving ownership.
Assuming the charging station uses a good random number generator and generates a new challenge for each session, this technique effectively prevents replay attacks.
However, the shortcoming with the current mechanism defined by ISO 15118 is the absence of any identifying credential of the charging station for the vehicle to check.
The vehicle does not include any additional information in the signed data, such as the current timestamp or information about the charging station it is connected to.
Thus, while the challenge prevents replay attacks, it does not effectively prevent relay attacks.
\changedtext{While Szak\'aly et al.\ \cite{szakaly_current_2025} already discussed the underlying TLS identification shortcomings that enable this attack, to the best of our knowledge, the plug and charge relay attack itself has not been previously described or implemented}.
\todo{[R13C] Section 5 describes the underlying issue in the TLS design that enables a relay attack against PnC, but it does not reference prior work that has already discussed this issue [21]. While previous work has not implemented the attack, it should nevertheless be acknowledged. The same applies to Section 7.1, which proposes adding geo-coordinates to TLS certificates. This has already been proposed by Szakaly et al. [21].}
\todoanswer{Added explicit acknowledgment of Szak\'aly et al.\ for discussing the underlying TLS issue. The same reference was also added to Section 7.1 for the geo-coordinates proposal.}

Normally, the use of TLS would prevent an attacker from creating a custom charging station emulator, as the lack of a valid certificate would prevent the victim vehicle from accepting it.
However, due to the certificate shortcomings described in Section \ref{sec:tls-problems}, an attacker could use certificates and private keys obtained from any compromised charging station to make the emulator impersonate any real charging station.
Since a certificate from any vendor is accepted by all vehicles, compromising a single poorly secured charging station is sufficient to attack vehicles at any other station, regardless of that station's own security posture.
This creates a weakest-link problem: the security of the entire ecosystem is bounded by the least secure charging station vendor.
With ISO 15118-20, which mandates mutual TLS authentication, the attacker vehicle must additionally present a valid vehicle TLS certificate when connecting to the \textit{regular charging station}.
This means the attacker must also have compromised at least one vehicle's TLS certificate and private key.
While this raises the bar compared to ISO 15118-2, \changedtext{payment handling is identical across both standard versions, so} the fundamental relay vulnerability remains: the plug and charge signature still contains no station-identifying information, and the attack proceeds in the same way once valid certificates for both sides are available.
\todo{[R13A] "With ISO 15118-20, which mandates mutual TLS authentication, the attacker vehicle must additionally present a valid vehicle TLS certificate when connecting to the regular charging station." Can't the attacker just use the certificate of their vehicle?}
\todoanswer{When the attacker modifies their own vehicle this is possible, but as described later, a more realistic approach is to use a man-in-the-middle device rather than modifying the vehicle itself.}
\todo{[R13B] The paper mentions that ISO 15118-20 introduces changes that affect the attack. This aspect should be discussed in more detail. At a minimum, a figure similar to Figure 1 should illustrate how the attack would proceed under ISO 15118-20 and whether these changes mitigate or alter the attack surface.}
\todoanswer{As was already mentioned above, the only difference is requiring a client certificate. The figure would not change between -2 and -20. We added an additional sentence to make this more clear.}

To perform such a relay attack, an attacker could build a charging station emulator and plug it into a victim's vehicle.
Simultaneously, the attacker vehicle is plugged into a regular charging station.
After initializing the charging communication with the victim's vehicle, the vehicle sends its charging contract certificate.
The attacker forwards this certificate to the charging station that the attacker's vehicle is plugged into, acting as if this certificate belongs to the attacker's vehicle.
The charging station then generates a random challenge and sends it to the attacker's vehicle.
Normally, the attacker would need the private key belonging to the victim's contract certificate in order to sign the challenge.
However, the attacker simply forwards the challenge to the fake charging station, which sends it to the victim vehicle.
The victim signs the challenge and returns the signature, which the attacker relays back to the regular charging station.
Since the signed data contains no station-specific information, there is no way for the regular charging station to detect that this signature was originally created by a different vehicle for a different charging station.
Figure \ref{fig:pnc-relay} visualizes the communication flow of this relay attack.

\begin{figure}[ht]
    \centering
    \includesvg[width=0.98\linewidth]{graphics/iso15118-relay.drawio.svg} 
    \caption{Plug and Charge Relay Attack}
    \label{fig:pnc-relay}
\end{figure}
\todo{[R13A] Figure 1 is somehow misleading. It seems that only 3 entities are involved from the chart, while there are 4 entities in the first line.}
\todoanswer{Modified the figure (now figure 2), adding a second line to visualize the breakup of the two ISO15118 communication sessions}


\section{Implementing and Combining Vulnerabilities for a Payment Fraud Proof of Concept} \label{sec:poc-impl}
\begin{figure*}
    \centering
    \includegraphics[width=0.9\linewidth]{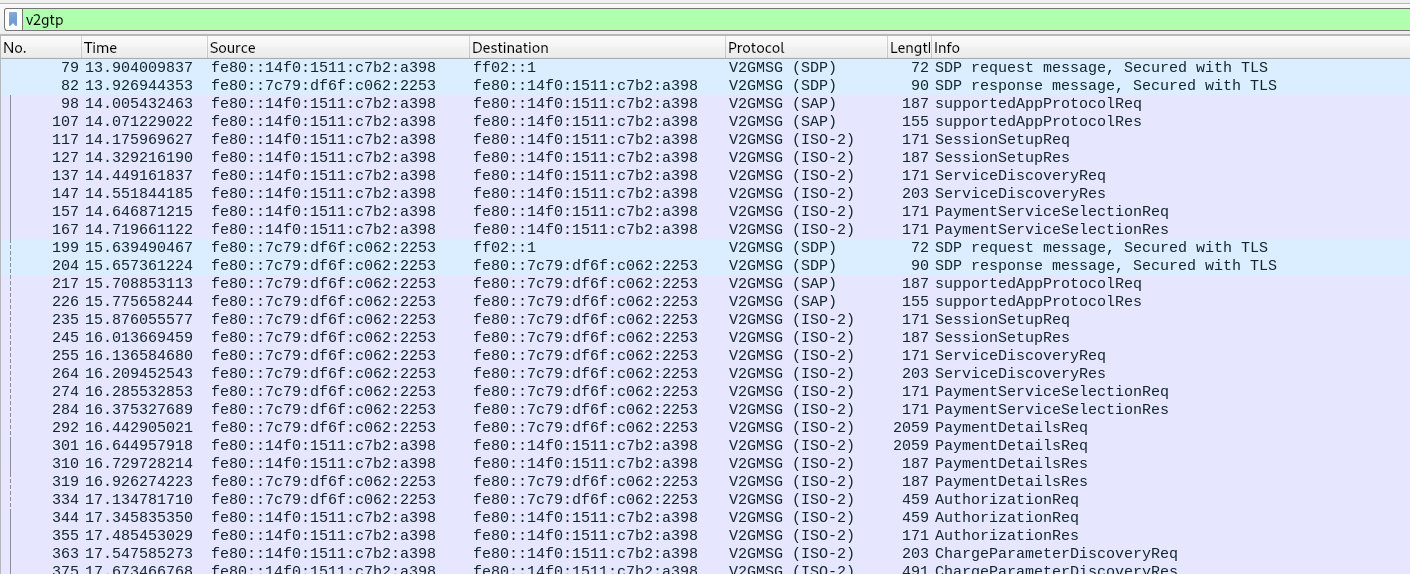}
    \caption{Wireshark Log of the two charging sessions during the attack}
    \label{fig:poc-wireshark-log}
\end{figure*}

The previous section described a plug and charge relay vulnerability allowing an attacker to charge their vehicle while making a second vehicle, the victim vehicle, pay for the charging session.
By combining this vulnerability with the vulnerability described in section \ref{sec:tls-problems}, this section describes a possible proof of concept proving the feasibility of the attack.
The software stacks, published as open source alongside this paper, can run entirely in a virtual environment, communicating with each other using local network interfaces.
Additionally, the proof of concept was successfully tested using real hardware: our own charging station as the \textit{regular charging station}, our EV simulator as the \textit{victim vehicle}, and two powerline modems acting as \textit{attacker vehicle} and \textit{fake charging station}.

\todo{[R13A] The setup is unclear. How is the fake charging station built? It seems that it should be connected to the victim vehicle while being able to send data to the regular charging station. It is somewhat explained in Section 6.1, but it should probably be part of the threat model, especially because it requires a modification on the attacker's side as well, which makes the "equipment supplied" attacker almost infeasible or, at least, more complicated.}
\todoanswer{The "unclear" points are exactly what is described in sections 6.2 and 6.3. First 6.2 describes how the data could be transmitted wirelessly between fake charging station and attacker vehicle. Section 6.3 mentions how a man-in-the-middle device could be used instead of modifying the attacker vehicle directly to make the attack more feasible.}

\subsection{Communication Participants and Requirements}
For this proof of concept, a custom PKI, as well as a set of virtual vehicles and virtual charging stations, is created.
The following lists the created communication participants, which are also present in figure \ref{fig:pnc-relay}:
\begin{itemize}
\item The \textit{victim vehicle}, which in reality could be any vehicle found parked on the street\changedtext{, after forcefully opening its charge port}.
\todo{[R13A] Is it true that it is sufficient to use "The victim vehicle, which in reality could be any vehicle found parked on the street"? Isn't the charging port locked and requires the keyfob to be opened? For Tesla, it is definitely locked, even though there may be vulnerabilities in how it could be opened. If true for the majority of vehicles, this significantly narrows the threat model and should be discussed.

[R13C] The attack is straightforward, but it should be noted that it might not be easy to trick the victim into using the malicious charging station controlled by the attacker. While I acknowledge that this depends on the situation, if the victim charges at a charging park with multiple DC fast chargers that all look the same (e.g. Tesla), they might get suspicious if there is one charger that looks completely different. Alternatively, the attacker would need to put in considerable effort to replicate a legitimate-looking charging station. To strengthen this point, I would recommend explicitly mentioning AC chargers, which are expected to also use ISO 15118-2/-20 in the near future.}
\todoanswer{Tesla is indeed a prime example, because there is public tooling and documentation on how to wirelessly open the electric charge port of a locked vehicle (e.g. using a flipper zero). For other vehicles opening the charge port via force is an option. We added a mention of this above.}
\item The \textit{attacker vehicle}. The goal of the attack is to charge the \textit{attacker vehicle} with energy being billed to the \textit{victim vehicle}.
\item The \textit{regular charging station}. This is the charging station that will be tricked into charging the \textit{attacker vehicle} while billing the \textit{victim vehicle} for the provided energy.
\item The \textit{compromised charging station} does not actually participate in the communication during the attack, but the attacker was previously able to extract this charging station's certificate and private key.
\item The \textit{fake charging station} is a purpose-built device. It does not need to actually provide power but merely needs to communicate with the \textit{victim vehicle} to perform the plug and charge handshake.
\end{itemize}

In order to create the \textit{attacker vehicle} as well as the \textit{fake charging station}, a customized ISO 15118 stack implementation is required.
Normally, when sending a \texttt{PaymentDetailsReq}, the vehicle includes its own certificate from its certificate store.
For the attack to be successful, the \textit{attacker vehicle} instead needs to send the certificate of the \textit{victim vehicle}, which was received by the \textit{fake charging station}.
Similarly, when the \textit{fake charging station} sends the \texttt{PaymentDetailsRes} to the \textit{victim vehicle} the included challenge needs to be identical to the original challenge received by the \textit{attacker vehicle} from the \textit{regular charging station}. 
Additionally, the \textit{attacker vehicle} will need to include the signature received by the \textit{fake charging station} from the \textit{victim vehicle} when sending its \texttt{AuthorizationReq} to the \textit{regular charging station}.
Therefore, a real-time communication between the \textit{fake charging station} and the \textit{attacker vehicle} is required, allowing them to exchange the certificate, challenge, and signature.

\subsection{Proof of Concept Implementation Details}
In reality, this communication is likely to be handled via a cellular network, allowing the \textit{fake charging station} to be at a different location than the \textit{attacker vehicle}.
In the proof of concept, the \textit{attacker vehicle} and \textit{fake charging station} exchange the certificate, challenge, and signature via a TCP connection, but run on the same machine.
If the \textit{attacker vehicle} reaches the step where it should send a \texttt{PaymentDetailsReq} before the certificate is available from the \textit{fake charging station}, it simply stalls and waits for it to arrive over the network.
For \texttt{PaymentDetailsReq} and \texttt{AuthorizationReq}, ISO 15118 specifies a response timeout of 2\,s \cite{isoiec_isoiec_2012}.
Since the relay only needs to forward small application-layer messages, a cellular network with typical round-trip times well below 100\,ms would provide more than enough margin for the relay to complete within the protocol's timeout constraints.

Implementation-wise, the proof of concept is based on the ISO 15118 stack formerly developed and published by SwitchEV, now EcoG \cite{ecog_iso_nodate}.
EcoG is a company selling commercial communication controllers and software for charging stations.
While the software stack is published as open source, allowing for easy modifications and attack implementation, it is also deployed in real charging stations.
In addition to the required changes for handling \texttt{PaymentDetailsReq}, \texttt{PaymentDetailsRes} and \texttt{AuthorizationReq} packets, the customized software stack also creates SSL key logs, allowing for easy inspection of encrypted traffic in Wireshark \cite{wireshark_wireshark_nodate}.
Our modified software stack is published under the original software license, allowing others to easily reproduce the attacks or try them against their own ISO 15118 implementations:
\anonymized{\url{https://anonymous.4open.science/r/plug-and-charge-relay-poc-325F/README.md}}{\url{https://github.com/securityinmobility/plug-and-charge-relay-poc}}

In order to run the proof of concept, in addition to the customized software stacks used for the \textit{fake charging station} and \textit{attacker vehicle}, two more ISO 15118 instances are required for the \textit{regular charging station} and \textit{victim vehicle}.
In a purely virtual setup, the same ISO 15118 stack can be used in its original unmodified version.
For our hardware-based test, a real charging station serves as the \textit{regular charging station} and our EV simulator acts as the \textit{victim vehicle}, while two powerline modems running the customized stack act as \textit{attacker vehicle} and \textit{fake charging station}.
Since the relevant vulnerabilities described in section \ref{sec:tls-problems} and section \ref{sec:pnc-vuln} are not specific to one implementation, but rather apply to all implementations of the standard, in theory, any compliant ISO 15118 implementation could be used.

\subsection{Proof of Concept Evaluation}
\todo{[R13C] Finally, while I appreciate that the attack has been implemented as a PoC and the source code has been made publicly available, I would have liked to see a more detailed evaluation of the attack in terms of quantitative metrics, for instance, the average overhead and delay introduced by the attack.}
\todoanswer{Section 6.2 already mentions an allowed delay of 2 seconds, which is why we did not further pursue timing measurements. Additionally the timing mitigation section mentions why response timing can be diverse anyways.}
Running the attack proof-of-concept produces a traffic log similar to the one shown in figure \ref{fig:poc-wireshark-log}.
In this screenshot, the IPv6 address of the \textit{attacker vehicle} ends on \texttt{a398}, while the IPv6 address of the \textit{victim vehicle} ends on \texttt{2253}.
At first, the \textit{attacker vehicle} initiates a charging session with the \textit{regular charging station}, but stalls after the \texttt{PaymentServiceSelectionRes}.
Normally, the \textit{attacker vehicle} would follow up with a \texttt{PaymentDetailsReq}, but it has to wait for the content from the \textit{victim vehicle} to relay the message.
The charging session between \textit{victim vehicle} and \textit{fake charging station} is started slightly later.
As soon as the \textit{fake charging station} receives the \texttt{PaymentDetailsReq} the included certificate is also used by the \textit{attacker vehicle} for its packet.
Similarly, as soon as the \textit{attacker vehicle} receives the \texttt{PaymentDetailsRes} including the challenge required to sign, the \textit{fake charging station} sends exactly this challenge to the \textit{victim vehicle}.
After the signed challenge is received from the \textit{victim vehicle} the communication with the victim stops.
Only the charging session with the \textit{attacker vehicle} continues.

\subsection{From Proof of Concept to the Real World} \label{sec:poc-real-world}
\todo{[R13A] It is not clear whether the real-world/hardware implementation is just theoretical or has actually been implemented.}
\todoanswer{Clarified at the start of this subsection that the hardware-based test described in Section 6.2 was actually carried out, and that this subsection discusses the additional engineering required to extend the attack to unmodified production vehicles.}
\todo{[R13C] The real-world practicality of the attack is not clear to me. The paper states that the attacker vehicle uses a customized ISO 15118 stack. It remains unclear how this would work for an off-the-shelf vehicle and whether this aligns with the threat and attacker model described earlier in the paper. Another difficulty is that the victim and attacker vehicle need to be synchronized at the beginning. The regular charging station will eventually close the connection with the attacker vehicle and switch state via basic signaling if the higher-layer communication has not progressed. In other words, the attacker vehicle would need to know when the victim plugs into the fake charging station to trigger the initialization of the charging session. As also acknowledged in Section 6.4, the real-world implementation is much more challenging and sophisticated.}
As described in Section \ref{sec:adversary-model}, the \textit{freeloader} and the \textit{contract-sharer} are the two adversaries which could directly benefit from abusing the described plug and charge relay vulnerability.

\changedtext{We successfully executed the attack both in a purely virtual setup and on real powerline hardware, as described in Section~\ref{sec:poc-impl}.
In this subsection, we discuss the additional engineering that would be required to extend the attack from our controlled hardware setup to unmodified, off-the-shelf production vehicles encountered in the wild.}
While our hardware-based test already demonstrates the attack using real charging infrastructure, a real-world scenario requires targeting actual vehicles rather than an EV simulator.
Building the \textit{fake charging station} which performs high-level communication with the \textit{victim vehicle} can easily be done using openly available charging communication evaluation kits \cite{chargebyte_evacharge_nodate, low_draindead_2025}.
Modifying the charging communication stack inside a vehicle, to build the \textit{attacker vehicle}, however, is a more complicated task.
A simpler approach would be to utilize known powerline communication vulnerabilities to design a man-in-the-middle device, which modifies the communication between an unmodified \textit{attacker vehicle} and a \textit{regular charging station}.
Previous work has demonstrated wireless powerline interception \cite{kohler_brokenwire_2023}, network infiltration via SLAC sniffing \cite{eder_charging_2025}, and man-in-the-middle attacks on charging communication \cite{dudek-v2g-2019, szakaly_dception_2026}.
Combining these techniques allows designing a wireless charging communication man-in-the-middle device.
This device could then perform plug and charge authentication with the charging station on one side and behave like a free charging station towards the \textit{attacker vehicle}.
Similar to keyless go relaying hardware utilized by car thieves, these known lower-layer vulnerabilities could be combined with the relay vulnerability presented in this paper to develop a pair of devices able to perform plug and charge relay attacks.
As described in Section \ref{sec:adversary-model}, this development and packaging of attack equipment into a consumer-ready product is the role of the \textit{equipment supplier}, whose existence would make the attack accessible to freeloaders and contract-sharers without any technical expertise.


\section{Possible Mitigations} \label{sec:mitigations}
\todo{[R13A] Except for Mitigation 7.3, which seems to be new and applicable, 7.2 is okay but comes from the literature and has the cited limitations, 7.1 will just reduce the scale of the attack but does not prevent it (also, GPS spoofing is cheap nowadays), and 7.4 concerns completely changing the employed system, which is quite unreasonable. Moreover, the proposed mitigations are described only theoretically, without any experiments to demonstrate their feasibility or to measure additional delays or performance degradation.}
\todoanswer{Since R13A is the shepherd: Would you prefer us to remove section 7.4? We still think a simpler payment system which works utilizing well established trust relations is a good approach, however we see the point of this not fitting 100\% into this paper, but rather a separate paper focusing on this new approach.}
This section discusses approaches for mitigating the plug and charge relay vulnerability presented in Section \ref{sec:pnc-vuln}, as well as an alternative payment mechanism that avoids the vulnerability entirely.

\subsection{Improving TLS Certificates for Charging Communication}
\todo{[R13C] Section 5 describes the underlying issue in the TLS design that enables a relay attack against PnC, but it does not reference prior work that has already discussed this issue [21]. While previous work has not implemented the attack, it should nevertheless be acknowledged. The same applies to Section 7.1, which proposes adding geo-coordinates to TLS certificates. This has already been proposed by Szakaly et al. [21].}
\todoanswer{Credit to Szak\'aly et al.\ for the geo-coordinates (and OCSP) proposal has been added below.}
For the plug and charge relay to work, the attacker needs to be able to impersonate a valid charging station towards the victim vehicle.
Due to the TLS weakness described in Section~\ref{sec:tls-problems}, a single compromised charging station certificate is sufficient to impersonate any station worldwide.
Fixing this does not directly address the relay vulnerability itself, but it would prevent this attack and potential future vulnerabilities from being exploited at scale, as the ability to impersonate arbitrary charging stations is a prerequisite for most attack scenarios.
\changedtext{Szak\'aly et al.\ \cite{szakaly_current_2025} already proposed two possible remedies for the TLS vulnerability: stricter OCSP checking and encoding geo-coordinates into charging station TLS certificates.}

\changedtext{OCSP provides a mechanism to revoke compromised certificates, which would limit the window during which an attacker can abuse a stolen charging station certificate.
However, OCSP checking is not mandatory for TLS leaf certificates in the plug and charge specification, and vehicles are not required to perform it.
Making OCSP verification mandatory for vehicles would ensure that revoked charging station certificates can no longer be used for impersonation, provided that compromises are detected and reported in a timely manner.}

\changedtext{Another way to strengthen the TLS trust model would be to encode geo-coordinates into charging station TLS certificates.}
Vehicles can then check the distance between their last GNSS position and the coordinates in the certificate.
When an attacker obtains a compromised certificate, it would only allow impersonation of stations near the compromised station's location, rather than any station worldwide.
This mitigation could be carried out as a TLS certificate extension, allowing backwards compatibility with vehicles that do not yet support the verification of geo-coordinates.

\subsection{Timing-Based Detection}
Distance bounding protocols that detect relay attacks through round-trip time measurements are well established in domains such as NFC and passive keyless entry.
Conti et al. propose a similar approach as a countermeasure to EVExchange \cite{conti_evexchange_2022}, measuring the latency of rapid challenge-response exchanges over the charging cable to detect the additional delay introduced by a relay.
In principle, such timing analysis could also be applied to detect the plug and charge relay presented in this paper.
However, production EV and EVSE controllers vary widely in processing power and system load, and the PLC channel over the charging cable is subject to variable latency from electrical noise and cable quality.
This makes it difficult to establish reliable timing thresholds that distinguish relay-induced delay from normal variation across the diverse ecosystem of charging hardware.

\subsection{Context-Binding the Plug and Charge Signature}
The relay attack succeeds because the plug and charge signature covers only the random challenge issued by the charging station.
Since the signed data contains no information about which charging station the vehicle believes it is connected to, the signature is valid at any charging station, enabling the relay.

A direct countermeasure is to bind the signature to the context of the current session by including the charging station's EVSE ID in the signed data.
During the TLS handshake, the charging station presents its certificate, which contains the EVSE ID as the certificate subject.
The vehicle can extract this identifier and concatenate it with the challenge bytes before computing the signature.
On the charging station side, the station concatenates its own EVSE ID with the challenge it issued and verifies the signature against the result.

This defeats the relay attack as follows:
When the \textit{victim vehicle} signs the challenge, it includes the EVSE ID of the \textit{fake charging station} as presented during the TLS handshake.
When the \textit{attacker vehicle} forwards this signature to the \textit{regular charging station}, the station verifies the signature against its own EVSE ID concatenated with the challenge.
Since the two EVSE IDs differ, the signature verification fails and the authentication is rejected.
The attacker cannot correct this mismatch without access to the victim's private contract certificate key.

This mitigation can be implemented in a backwards-compatible manner without changing the message format defined in ISO~15118.
The challenge issued by the charging station is 16 bytes long.
Instead of filling these bytes entirely with random data, an updated charging station embeds its country code and operator identifier (5 bytes) within the challenge, using the remaining 11 bytes for random data.
Updated vehicles extract the EVSE ID from the charging station's TLS certificate and verify that the corresponding country and operator bytes in the challenge match; if they do not, the vehicle rejects the session, defeating the relay attack.

This approach is backwards compatible in both directions:
vehicles that have not been updated treat the challenge as opaque random data and sign it as before, so they continue to function with updated charging stations.
Conversely, updated vehicles connecting to a legacy charging station that issues fully random challenges can detect the absence of the expected EVSE identifier and reject the session.
As a result, updated vehicles are protected regardless of whether the charging station has been updated, while legacy vehicles retain full functionality with both legacy and updated infrastructure.

\subsection{Replacing Plug and Charge with Out-of-Band Payment}
An alternative to fixing the plug and charge mechanism is to bypass it entirely.
After establishing a TLS connection with the charging station, the vehicle forwards the EVSE ID to a manufacturer-provided backend, which initiates the charging session via the OCPP \texttt{RemoteStartTransaction} command — the same mechanism already used by mobile charging apps today.
This eliminates the need for the cross-vendor PKI required by plug and charge, and requires no changes to ISO~15118, OCPP, or the charging station software.
Additionally, this approach would greatly reduce complexity, as it removes the need for enrolling contract certificates and designing custom challenge-response procedures for plug and charge.

This approach has two drawbacks:
it makes the vehicle manufacturer a gatekeeper for payment, and it requires an internet connection, unlike plug and charge which can operate offline.
However, ISO~15118 already gives manufacturers control over which contract certificates are enrolled, and the trend towards connected charging infrastructure — reinforced by regulations such as the EU Alternative Fuels Infrastructure Regulation \cite{noauthor_regulation_2023} — makes the offline limitation increasingly marginal.
\changedtext{A full evaluation of this alternative payment approach, including its security properties, deployment considerations, and impact on existing roaming agreements, is beyond the scope of this paper and should be explored in more comprehensive studies in future work.}


\section{Conclusion}
In this paper, we presented a novel relay attack against the plug and charge payment mechanism of ISO 15118.
The attack exploits the absence of station-identifying information in the plug and charge signature, combined with the inability of vehicles to verify charging station identity through TLS certificates.
As shown in our proof-of-concept implementation, these weaknesses make it possible to build a relay device, similar to keyless go relaying, that allows adversaries to charge a vehicle while a different vehicle is billed for the charged energy.

We discussed direct mitigations for the relay vulnerability, including binding the plug and charge signature to the charging station identifier and timing-based detection of relay delays.
Additionally, we proposed an alternative out-of-band payment approach that shifts payment handling to the vehicle manufacturer's backend, reusing existing remote start mechanisms.
This approach removes the need for the complex cross-vendor PKI required by plug and charge and can be implemented without changes to ISO 15118 or charging station software.

As plug and charge adoption grows, addressing this vulnerability is critical before it becomes widely exploitable.

\section{Ethical Considerations}
\todo{[R13B] The authors should provide more details on any response received from the ISO 15118 committee, as this would help understand the practical impact and disclosure process.}
\todoanswer{Sadly no response from CharIN so far. We still hope to receive one before the camera ready deadline in order to update this section. We are in contact with some of the major charging station manufacturers. Added a mention of this below.}
We disclosed the plug and charge relay vulnerability to CharIN, the organization managing the Combined Charging System and the ISO 15118 standard, prior to publication.
\changedtext{In addition, we contacted several major charging station manufacturers directly; while they acknowledged the disclosure, they did not consider immediate mitigation necessary on their side.}
As plug and charge is not widely deployed today \cite{szakaly_current_2025, low_fast_2024}, the immediate risk to users is limited.
Furthermore, performing the attack in practice requires obtaining a valid TLS certificate trusted by the victim vehicle, which requires exploiting and extracting one from a real-world charging station, potentially using a zero-day exploit.

\section{Open Science}
Our proof-of-concept implementation is publicly available at:
\anonymized{\url{https://anonymous.4open.science/r/plug-and-charge-relay-poc-325F/README.md}}{\url{https://github.com/securityinmobility/plug-and-charge-relay-poc}}
Apart from the modified charging communication stack, this repository also includes network dumps of successful demonstrations of the plug and charge payment fraud attack.
By using the free and open source dissector for charging communication \enquote{dsV2Gshark} \cite{dspace_dsv2gshark_nodate}, these files can be viewed using Wireshark \cite{wireshark_wireshark_nodate}.

\section*{Acknowledgment}
\anonymized{\textit{redacted for anonymized version}}{
This work was created in the research project \enquote{Elektromobiles Sicheres Laden} (ESiLa) funded by the Bavarian Ministry of Economic Affairs, Regional Development and Energy under grant DIK0512/01.
}


\bibliographystyle{plain}
\bibliography{references}

\end{document}